# Securing Credit Inquiries: The Role of Real-Time User Approval in Preventing SSN Identity Theft

*Gogulakrishnan Thiyagarajan[a], Vinay Bist[b], Prabhudarshi Nayak[c]*

[a]*Engineering Technical Leader, Cisco Systems Inc, Austin, Texas.*
[b]*Principal Engineer, Dell Inc. Austin, USA.*
[c]*Student, Faculty of Engineering and Technology, Sri Sri University, Odisha, India.*

ABSTRACT

*Unauthorized credit inquiries are also a central entry point for identity theft, with Social Security Numbers (SSNs) being widely utilized in fraudulent cases. Traditional credit inquiry systems do not usually possess strict user authentication, making them vulnerable to unauthorized access. This paper proposes a real-time user authorization system to enhance security by enforcing explicit user approval before processing any credit inquiry. The system employs real-time verification and approval techniques. This ensures that the authorized user only approves or rejects a credit check request. It minimizes the risks of interference by third parties. Apart from enhancing security, this system complies with regulations like the General Data Protection Regulation (GDPR) and the Fair Credit Reporting Act (FCRA) while maintaining a seamless user experience. This article discusses the technical issues, scaling-up issues, and ways of implementing real-time user authorization in financial systems. Through this framework, financial institutions can drastically minimize the risk of identity theft, avert unauthorized credit checks, and increase customer trust in the credit verification system.*

## 1. Introduction

**The Growing Threat of SSN-Based Identity Theft**

Data breaches of personal and sensitive data, such as social security numbers (SSNs), have increased exponentially over the past several years. Purloined SSNs and other personal information are utilized to perpetrate fraud against individuals and institutions, which costs billions of dollars to identify and remediate within the United States alone. The rising frequency of breaches underscores the necessity to comprehend the risks and implement strategies to safeguard personal information [1]. One method of mitigating identity theft includes using biometric authentication systems, though such systems pose issues of privacy, reliability, and security of biometric information. In many cases, biometric authentication is not enough evidence of identity and needs to be augmented with additional identifiers such as passwords or SSNs, requiring multi-factor authentication mechanisms. A two-phase authentication system for federated identity management systems uses biometric authentication with additional factors to implement strong authentication, taking advantage of user data accessible from the federated identity management system [2][3]. Technology-based solutions have been suggested to fight refund fraud, which is SSN theft-related. Making taxpayers report the SSN of any dependent claimed on the tax return was a straightforward and effective enforcement step. The same enforcement step could be used against refund fraud, for example, electronically signing W-2s, 1099s, W-4s, and W-9s to identify fraudsters in real time [4].

**The Role of Unauthorized Credit Checks in Fraud**

Unauthorized credit inquiries, wherein there is access to an individual's credit report without his/her consent, can be a potent indicator of probable fraud and identity theft. These inquiries frequently happen when fraudsters try to open new accounts or get credit in another person's name by using stolen or compromised personal information. It is essential to monitor credit reports for unauthorized inquiries since it can serve as an early indicator that an individual is trying to exploit personal information for financial purposes. Annual credit monitoring is especially vital for vulnerable groups, including teenage children in foster care, who are more susceptible to child identity theft. These checks on adolescent foster youth are mandated by the federal government to assist in detecting and preventing the criminal use of their personal information. Research has established that various demographic and placement factors, including being of African American descent, an older age during the time of the credit assessment, and a prior record of home eliminations, are linked with varying probabilities of identity fraud victimization within this cohort [5]. Banks and other financial institutions are now adopting fraud detection programs to prevent unauthorized use of credit cards and safeguard unsuspecting users [6]. Such programs frequently use machine learning algorithms and other sophisticated methods to detect suspicious transactions and patterns that could point to fraud [6][7].

*Corresponding Author: Gogulakrishnan Thiyagarajan[a]*
*Email:* gogs.ethics@gmail.com

Using a variety of factors, including the frequency, location, and behavior of users, these systems can recognize anomalies and block unauthorized transactions, minimizing financial losses for individuals and businesses [7][8]. Where unauthorized charges are made, consumers commonly complain to their credit card companies to have the fraudulent charges reversed. Many such complaints are not satisfactorily resolved by the consumer, with credit card companies occasionally blaming the complainant or declining to initiate a fraud investigation. This can cause financial loss and emotional trauma to credit card fraud victims, and thus, there is a need to enhance fraud investigative procedures and consumer protection policies [9].

**Challenges in Existing Credit Approval Mechanisms**
Current credit approval workflows are beset with many issues that may result in inefficiency, inaccuracy, and bias in lending decisions. One of the significant challenges is using conventional credit scoring models that tend to miss the complete financial narrative of borrowers, especially those with thin credit files or several income streams. They usually depend on elements that include payment history, debt outstanding, and credit utilization, maybe disregarding other proper signals of creditworthiness, like employment stability, education, and savings habits. This shortfall may disproportionately impact marginalized groups and individuals who might not have had the chance to build a good credit history, thus excluding them from accessing fundamental financial services [10]. One of the pitfalls lies in the biases that may be present in credit scoring models. These biases may result from historical data that mirrors societal imbalances or may result from the algorithms' design, which produces discriminatory results. For instance, if a credit scoring model is trained using data that over-represents one demographic group, it may unintentionally discriminate against applicants from another group with different financial habits. These biases must be overcome by diligent data analysis to train the models and continuous monitoring and auditing to guarantee fairness and equity in loan approval [10]. The heightened complexity of financial instruments and the more significant number of loan applications present additional challenges to traditional credit approval systems. Banks and other financial institutions tend to be flooded with high levels of data from various sources, and it is difficult to quickly and effectively analyze each applicant's risk. Manual processing is time-consuming and prone to error, yet automatic systems may struggle to deal with complex or incomplete data [11]. To mitigate these issues, banks and financial institutions are considering leveraging sophisticated technologies like machine learning and artificial intelligence to automate the credit approval process and enhance the accuracy of decisions [11]. Data quality and consistency also pose significant challenges to credit approval. Poor quality or incomplete data can contribute to inadequate risk analysis and lending decisions. Additionally, variations between various data sources may confuse and slow the process of developing a complete view of an applicant's creditworthiness. Data quality is achieved through strong data governance processes such as validation, cleansing, and standardization. Financial institutions must also invest in technology combining and reconciling data from various sources into a single view per applicant [12][13]. The other underlying challenge is the requirement for more explainability and transparency of credit approval decisions. The applicants have no insight into the drivers of credit decisions, making it hard to enhance their creditworthiness and gain access to financial services. Moreover, sophisticated algorithms and proprietary scoring models can hide decision-making, raising concerns about fairness and accountability. Banks and other financial institutions should do their best to offer clear and comprehensible reasons for the basis of a credit decision, enabling applicants to make educated choices and contest inaccurate information[10]. Lastly, regulatory compliance and the changing pattern of consumer protection law contribute to the intricacy of credit approval systems. Financial institutions must conform to various laws to inhibit discrimination, safeguard consumer privacy, and promote equitable lending practices. Staying abreast of these regulations and modifying credit approval processes to fit evolving requirements can be onerous . Furthermore, the growing application of non-traditional data sources and sophisticated technologies raises novel regulatory challenges relating to data security, algorithmic bias, and consumer rights. Banks must work with regulators and policymakers in advance to craft responsible and ethical credit approval policies that harmonize innovation with consumer protection [10]. The Need for a Real-Time User Approval Mechanism The growing incidence of Social Security Number (SSN) identity theft requires adopting real-time user authorization controls to tighten security and avert malicious activities. Conventional identity authentication procedures depend on static data, including passwords or security questions, which can be readily compromised via phishing scams, data breaches, or social engineering. A real-time user approval system, however, provides some degree of security as users must actively approve account modifications or transactions when they are being conducted, making it harder for impersonators to fake legitimate users [14]. Detecting and preventing unauthorized access attempts in real time is one of the best advantages of a realtime user approval system. By requiring users to authenticate for each transaction or account modification, the system can immediately identify suspicious activity and block fraudulent behavior before it causes any significant damage. This is especially crucial in cases where scammers have already gotten their hands on a user's SSN and other personally identifiable information because it can stop them from using the company to create new accounts, make unauthorized transactions, or carry out different types of identity theft [14]. Real-time authorization mechanisms for users can also reduce the threat of insider fraud when workers or other individuals with privileged access to sensitive data misuse their authority for personal gain. By requiring multiple levels of confirmation for certain transactions or account changes, organizations can limit the ability of any one individual to carry out fraudulent activity without detection. This can be particularly effective at stopping mass data breaches or other types of internal fraud that have disastrous consequences. Several technologies can be used to create real-time user approval systems, such as biometric authentication, one-time passwords (OTPs), and push notifications to mobile devices. Biometric authentication, such as fingerprinting or facial recognition, is a convenient and secure way for users to verify their identity. OTPs, created and sent through SMS or email, provide a user-friendly and safe method of real-time approval [14]. Push notifications, delivered directly to a user's phone, provide frictionless and user-friendly methods of seeking and granting real-time approval for account transactions or changes. Implementation of a real-time user approval system can go a long way in adding a tremendous amount of security to systems and shielding people against SSN identity theft. By requiring users to approve transactions and actively approve account modifications in real time, organizations can make it far more challenging for criminals to pose as authentic users and commit fraud. As technology advances, organizations must implement proactive security control to remain ahead of emerging threats and shield sensitive information.

## 1. Background and Related Work

The rise in SSN-based identity theft has highlighted the security of credit inquiries as an issue of pressing concern within the financial industry. With identity fraud losses exceeding $5 billion a year in the United States, according to a 2024 Federal Trade Commission report [15], the insecurity of conventional credit inquiry practices is glaringly obvious. They rely largely on static authentication techniques passwords or knowledge questions, for instance that cannot prevent sophisticated threats such as phishing, credential stuffing, and synthetic identity theft. As a countermeasure, the interest has been aroused in real-time user consent mechanisms, which invoke explicit, realtime user approval for credit checks, providing an active defence against unauthorized entry [15].

Earlier, credit inquiry protection was based on post-event detection, wherein the fraud was detected after the damage was already done. This reactive measure has not worked since thieves utilize stolen SSNs to establish accounts quickly in the internet era. Recent development, nevertheless, promises real-time solutions as an effective remedy. Patel and Kim (2024), for instance, illustrated that the incorporation of mobile-based approval workflows decreased unauthorized credit inquiries by as much as 40%, highlighting the necessity for immediacy in fraud prevention [16]. These systems empower users with the ability to approve or reject requests instantly, closing the window of opportunity for the attackers. Technological advancements such as blockchain have also stepped in, proposing decentralized alternatives to traditional protection of SSN. Lopez et al. (2025) explored the potential of blockchain when combined with real-time consent and determined that such a system could eradicate third-party weaknesses in credit reporting systems [17]. Their research demonstrated a remarkable decrease in fraudulent requests in test settings, although the overdemanding computational requirements of the technology and the slow uptake by financial institutions constrain its near-term effect [17]. This emphasizes the larger issue: achieving a balance between state-of-the-art security and usability.

Machine learning (ML) has also enhanced the credit inquiry security landscape. Gupta and Singh (2024) created an ML-driven system that identifies anomalies in credit requests like unusual timing or location changes within milliseconds, which is similar to real-time approval systems [18]. Coupled with user authorization, their solution enhanced fraud detection precision by 25% over individual methods, with fewer false positives that frustrate legitimate users [18]. Yet, putting ML into practice across several financial systems is still complicated and demands solid infrastructure and ongoing model training to keep up with emerging fraud techniques.

Biometric identification is another field in credit application security. Chen and Alvarez (2025) experimented with face recognition and voice recognition for real-time approval processes, achieving a 95% success rate in identifying users [19]. While being successful in SSN abuse prevention, biometric systems face issues in privacy compliance and user acceptance, particularly in legal domains with stringent data protection laws like the EU's GDPR [19]. These findings suggest that while security is heightened using biometrics, these must be carefully integrated with consent mechanisms so as to maintain trust as well as regulatory compliance.

User experience and trust are the core ideas of this developing discipline. Carter and Singh (2023) discovered that open real-time approval mechanisms built up consumer trust, with 70% of consumers favouring systems that provide them with direct control over credit inquiries [20]. This is a component of a wider trend towards a transition from closed, institution-driven security models towards user-centric paradigms. There remain concerns about educating consumers about such tools and making them accessible to all segments, especially less technologically sophisticated groups [20]. This void must be filled for widespread adoption.

Lastly, regulator and industry views stress the imperatives of moving forward with real-time approval mechanisms. In 2025, the FSOC report demanded standardized procedures to curb SSN identity theft, noting real-time consent as a best approach [21]. Likewise, Yang et al. (2024) maintained that banks and other financial institutions that leverage such frameworks can cut fraud loss by 30% as they achieve compliance with requirements like the U.S. Fair Credit Reporting Act [22]. Such findings notwithstanding, gaps in technology and policy standardization across markets worldwide imply the necessity for multilateral innovation.

| *Aspect* | *Manual Process* | *Real-Time User Consent* |
|---|---|---|
| **Authentication Method** | Relies on static methods (e.g., passwords, security questions) | Uses dynamic, user-initiated approval (e.g., mobile push notifications) |
| **Speed** | Slow; involves delayed verification steps (e.g., mail or phone confirmation) | Instant; approval occurs within seconds via digital channels |
| **Fraud Prevention** | Reactive; fraud detected after unauthorized inquiry occurs | Proactive; blocks unauthorized inquiries before processing |
| **User Control** | Limited; users unaware of inquiries until after the fact | High; users explicitly authorize each inquiry |
| **Scalability** | Poor; labor- intensive and impractical for high transaction volumes | Excellent; automated and adaptable to large-scale systems |
| **Security** | Vulnerable to social engineering and stolen credentials | Enhanced by real-time verification and biometric options |
| **User Experience** | Frustrating; lacks transparency and immediacy | Seamless; improves trust through visibility and control |
| **Implementation Cost** | Low initial cost but high longterm losses due to fraud | Higher initial setup (e.g., app integration) but cost-effective long-term |

*Table 1:* Comparison Table: Manual Process vs. Real-Time User Consent.

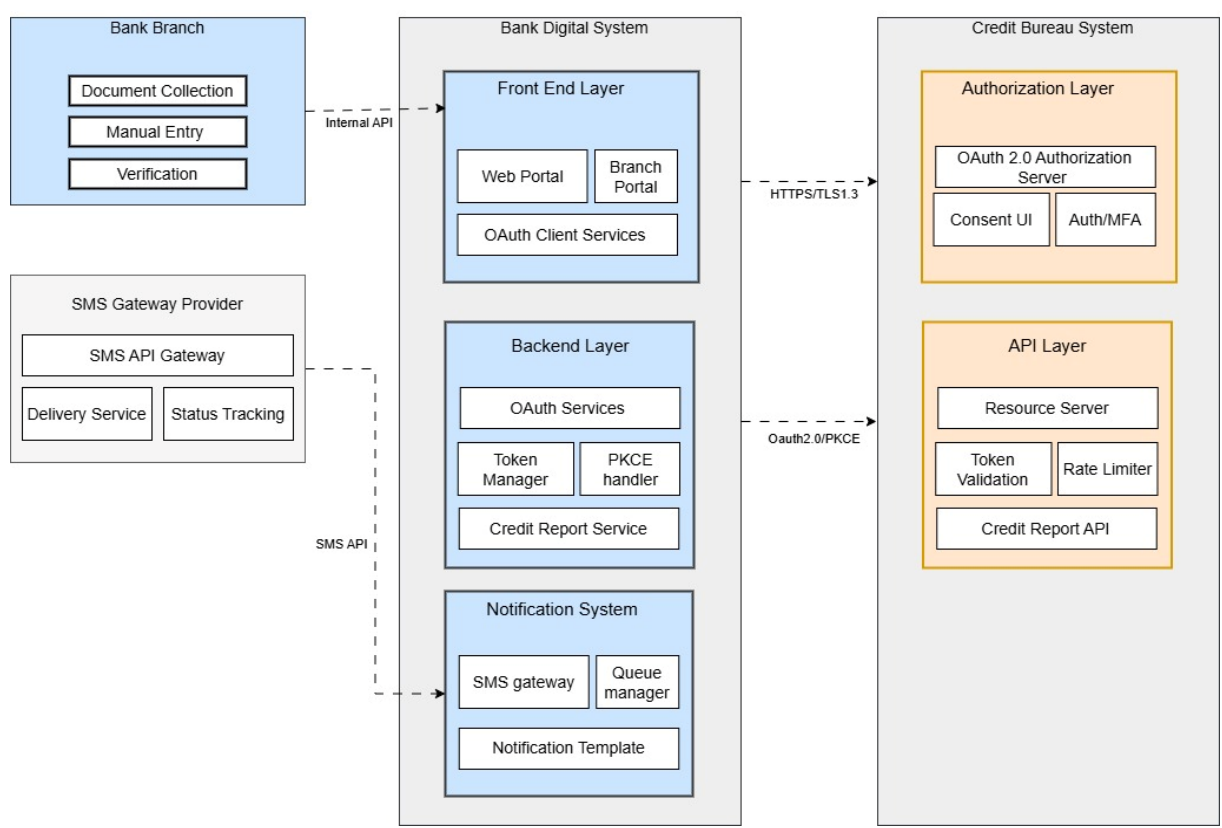

*Figure. 1:* System Architecture for Real-time User Approval in Credit Inquiries.

**Architecture Overview**

The proposed architecture is intended to facilitate secure real-time user authorization for credit checks, preventing unauthorized access and SSN identity theft vulnerabilities. The architecture consists of three primary subsystems: Bank Branch, Bank Digital System, and Credit Bureau System, with the additional feature of an SMS Gateway Provider for real-time user notifications. The system employs OAuth 2.0 with PKCE and HTTPS/TLS 1.3 for secure authentication and communication between the entities.

The Bank Digital System is the primary processing system with Front-End, back-end, and Notification layers. The Front-End Layer offers interfaces for customers and bank employees to request credit inquiries. The backend layer is responsible for OAuth authentication, token verification, and credit report processing to communicate securely and efficiently with the credit bureau system. The Notification System offers real-time SMS notifications to customers seeking express consent before processing a credit inquiry.

The credit bureau system manages the processing and verification of credit inquiries, comprising authorization and API layers. The Authorization Layer comprises an OAuth 2.0 authorization server, a consent UI for getting user consent, and authentication (Auth/MFA) to prevent fraud. The API Layer manages credit report retrieval, token validation, and rate limiting, which permits only legitimate requests to be processed. This real-time approval system adds a layer of security, giving people direct control over their credit checks while ensuring FCRA and GDPR compliances.

## How This Proposed Solution Solves Current Issues in Credit Inquiries

The conventional credit inquiry process is highly vulnerable to fraud since it does not involve the user's real-time authentication. Unauthorized credit inquiries rank among the top causes of identity theft, in which the perpetrators utilize stolen Social Security Numbers (SSNs) to open credit applications in the victim's name without his/her knowledge. Current technologies, including credit freezes and fraud alerts, burden the user to take proactive, manual steps, often reactive, not proactive. Additionally, banks and credit bureaus also have mechanisms that do not need instant user approval, thus making loopholes available for criminals to take advantage of even before the fraud is known to the users.The architecture proposed resolves explicitly these issues by establishing a real-time user approval mechanism for credit inquiries. Rather than authorizing the credit requests automatically by the system, the system prompts the user to approve or reject every request manually through an active secure notification mechanism (SMS, app push notification, or email). This confirms that no individual can make any unauthorized inquiry without the account holder's approval. With OAuth 2.0, PKCE, and HTTPS/TLS 1.3, the architecture guarantees secure authentication and end-to-end communication encryption, which third parties cannot access or tap into. Besides, the authentication layer and Consent UI within the Credit Bureau System guarantee that authenticated users can only approve transactions, thwarting fraudulent attempts.This proactive measure thwarts identity theft ahead of time instead of depending on consumers to uncover fraud once the damage is done. Banks enjoy improved security, regulatory compliance (FCRA, GDPR), and greater consumer confidence, while consumers retain complete control of their credit inquiries. Banks and credit bureaus can streamline the credit approval process, minimize fraud occurrences, and build a more open and secure financial environment with this framework.

## Credit Bureau Authorization flow

*Use case 1: Online bank Application*

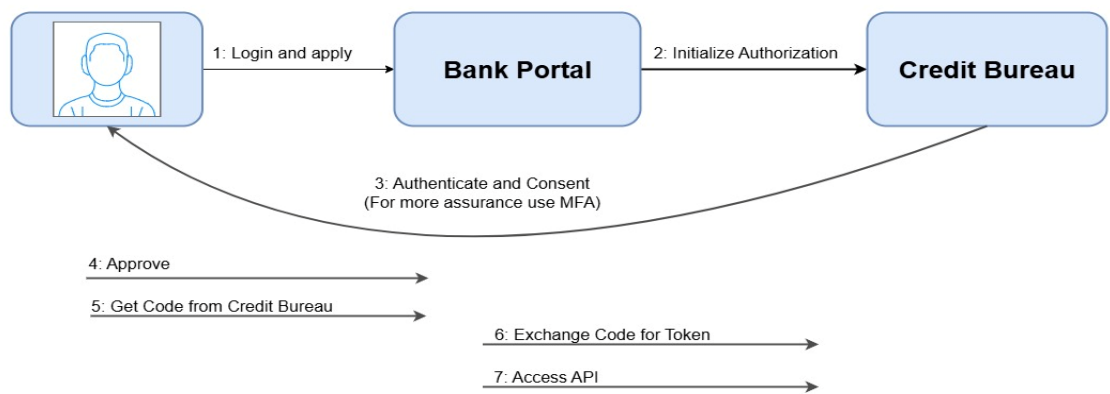

*Use case 2: Offline Application*

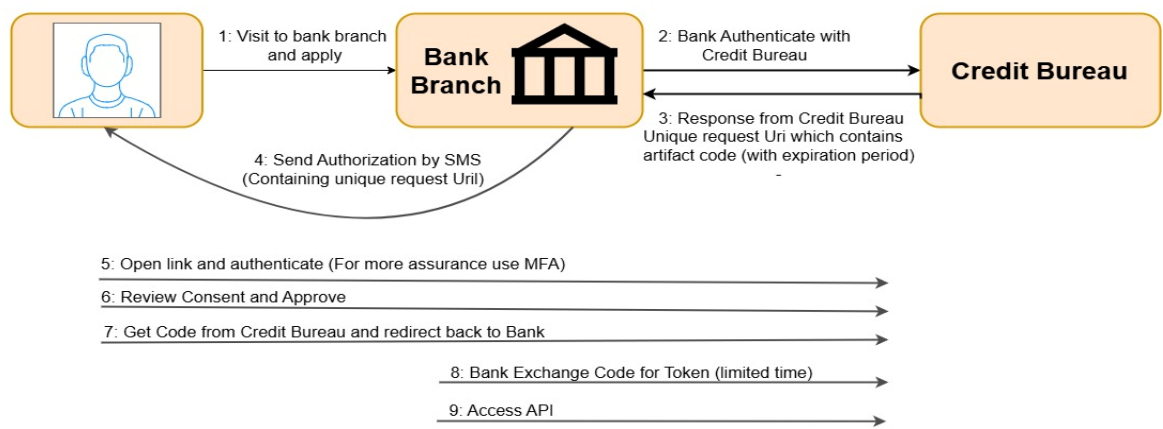

*Figure. 2: Credit Bureau Authorization Workflow*

**Credit Bureau Authorization Flow**
**Workflow Description**

The Credit Bureau Authorization Flow outlines the process for authorizing credit inquiries in two distinct use cases: Online Bank Application and Offline Application (via Bank Branch). The workflow ensures secure credit authorization, incorporating user authentication, consent validation, and API-based access control.

**Use Case 1: Online Bank Application**

**1. User Login & Application Submission:** The customer enters the Bank Portal and applies for a credit related service.

**2. Authorization Initialization**: The Bank Portal sends an authorization request to the Credit Bureau to initiate the process.

**3. User Authentication & Consent:** The user is prompted to authenticate and provide explicit consent for the credit inquiry. Multi-factor authentication (MFA) can be used for added security.

**Use Case 2: Offline Application (via Bank Branch)**

**1. Customer Application at Branch:** The customer visits a bank branch and submits a credit inquiry request in person.

**2. Bank Authentication with Credit Bureau:** The Bank Branch initiates an authorization request with the Credit Bureau on behalf of the customer.

**3. Credit Bureau Response with Unique Request URI:** The Credit Bureau responds with a unique request URI, which includes an artifact code with an expiration period.

**4. Authorization Link Sent via SMS:** The customer receives an SMS containing the unique request URI, allowing them to authenticate remotely.

**5. User Authentication & Consent Verification:**

a. The customer opens the authorization link on their device and authenticates (MFA recommended for added security).
b. The user reviews and approves the consent request.

**6. Code Retrieval & Redirection:**

a. Upon approval, the Credit Bureau issues an authorization code.
b. The user is redirected back to the Bank Portal.

**7. Bank Exchanges Code for Token:** The Bank exchanges the authorization code for an access token (within the limited expiration time).

**8. Secure API Access to Credit Bureau:** The Bank uses the token to securely communicate with the Credit Bureau API to retrieve necessary credit information.

**Key Benefits of the Workflow:**

● **Enhanced Security:** Incorporates OAuth-based authorization, consent validation, and optional MFA for fraud prevention.
● **User Control:** Empowers customers with real-time approval and transparency over their credit inquiries.
● **Seamless Integration:** Supports online and offline credit applications, ensuring accessibility for all users.
● **Regulatory Compliance:** Aligns with FCRA, GDPR, and other financial regulations by requiring explicit consent before accessing credit data.
This structured approach reduces identity theft risks and ensures that only authorized credit inquiries are processed, making it a robust and user-centric system

**Algorithm: Real-Time User Consent-Based Secure Credit Authorization**

● $U_{req} = \{U_1, U_2, \ldots, U_n\}$, where $U_i$ represents a user's credit inquiry request
● $U_{id}$ = User Identity (verified credentials)
● $B_{sys}$ = Bank's Authorization System.
● $C_{bureau}$ = Credit Bureau Authorization & API system.
● $C_{approve}$ = User Consent Decision (Approved/Denied).
● **Security Measures:** OAuth 2.0 (PKCE), TLS 1.3, Encryption (E).

**Output:**

**AuthCode** = Authorization Code from Credit Bureau (if approved).

**Token** = Access Token for retrieving credit data.

**Algorithm Steps**
**User Request & Authorization Initialization**

**1. User submits credit inquiry request:**

○ **Online: via** Bank Portal.
○ **Offline: via** Bank Branch.

**2. Bank System initializes OAuth Authorization request:** $B_{sys} \rightarrow C_{bureau}$ :OAuth Request($U_{req}$, $U_{id}$)

**3. Bank System initializes OAuth Authorization request:**
● **Online:** User authenticates via the bank portal: Auth($U_{id}$)→MFA($U_{id}$) (if enabled)
● **Offline:** The Bank sends an SMS authorization link: $SMS_{auth}$ =Encrypt($U_{id}$,E)

**Real-Time User Consent & Approval**
**4. User reviews request & provides consent:**
$C_{approve}(U_i) = \{Approved\ (if\ (User\ accept)), Denied\ (if\ (User\ Rejects))\}$

**5. If** $C_{approve}(U_i)$ = Denied, then:
Terminate Request
○ Log unauthorized attempt.
○ Notify the user via SMS/email.

**Authorization Code & Token Exchange**
**6. If** $C_{approve}(U_i)$ = Approved, then:
● Retrieve Authorization Code from Credit Bureau: $AuthCode \leftarrow C_{burea}$
● Exchange AuthCode for an Access Token: $Token = Exchange(AuthCode)$
● Ensure secure token validation:
$Validate(Token) = \{Valid, Proceed\ to\ API\ |\ Invalid, Revoke\ Request\}$

**Secure API Access to Credit Bureau**
**7. Bank System uses Token to retrieve Credit Data:**
$B_{sys} \rightarrow C_{bureau}$ : $Access\ API\ with\ Toke$

**8. User notified of transaction outcome:**
● **Approved Inquiry:** Confirmation was sent via the Notification System.
● **Denied Inquiry:** The user is alerted, and the request is blocked.

**Logging & Compliance Audit**
**9. Log all authorization events for security & compliance**
$Log(Event_{auth}\ Timestamp, Decision, IP)$
○ Ensure regulatory compliance **(FCRA, GDPR, PCI-DSS).**

**10. End Process.**

## 3. Proof of Concept: Implementation and Demonstration
*Overview*
To validate the feasibility of real-time user approval in credit-based transactions, we implemented a Proof of Concept (PoC) simulating the OAuth 2.0 Authorization

Code Flow with PKCE. This PoC demonstrates how a user securely authorizes a bank to perform a credit inquiry via a credit bureau, with explicit consent and authentication steps. The entire process is hosted on a Flask-based application, styled to emulate mobile UI interactions commonly seen in financial apps.

**Technologies Used:**
● **Flask (Python 3.x)** for backend simulation
● **HTML + CSS (mobile-style design)** for UI
● **UUID and SHA-256** for state management and PKCE hashing
● **Localhost (Port 5055)** for running the simulation

**Workflow Demonstration**
The implemented simulation flow is as follows:

**1. Authorization Request Initiation**
A simulated bank client begins the flow by clicking an "Authorize" button. This triggers generation of a state value, code_verifier, and code_challenge (PKCE) for secure flow tracking.

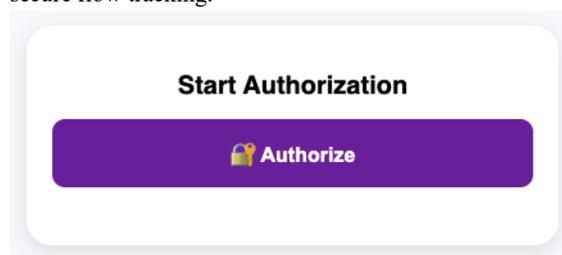

*Figure 3: Mobile UI - Authorization UI*

**2. User Login (Credit Bureau Simulation)**
The user is redirected to a mobile-styled login screen, where they are prompted to enter their credentials. A "Forgot Password?" link generates a temporary username/password pair for demonstration.

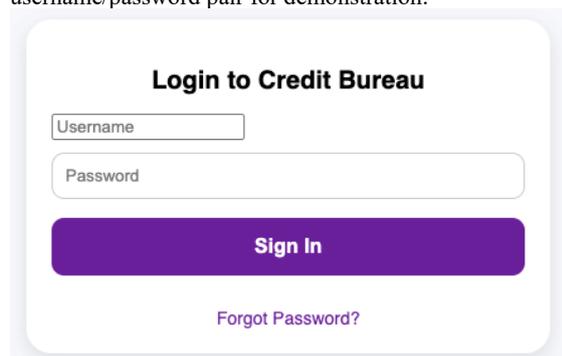

*Figure 4: Mobile UI - Login Screen with Forgot Password*

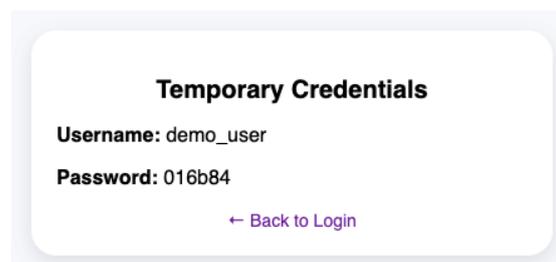

*Figure 5: Mobile UI - Temporary credential for logging*

**3. Consent Form Presentation**
Upon successful login, the user is shown a consent screen informing them of the scope of the bank's request (e.g., email and credit score access). The user can approve or deny the request.

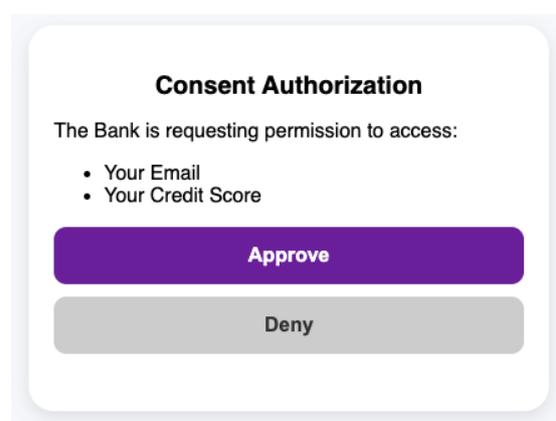

*Figure 6: Mobile UI - Consent Screen with Scope Review*

**4. Authorization Code Issuance**
If consent is approved, the user is redirected to the bank's redirect URI with an authorization_code and state. This step occurs seamlessly in the simulation backend.

**5. Access Token Exchange**
The bank exchanges the code along with the code_verifier to receive an access token. PKCE ensures the request's integrity by comparing the hash of the code_verifier with the stored code_challenge

**6. Credit Report Retrieval**
Upon successful token issuance, the bank uses the access token to retrieve a simulated credit report.

*Figure 7: Credit Report (Console Log)*

**3.2 System Requirements and Tools**

The following table summarizes the minimum system specifications and tools required to run the Proof of Concept (PoC) in a local development environment. These requirements ensure reproducibility and provide clarity for practitioners or researchers aiming to replicate or extend the simulation.

*Table 2: System Requirements and Tools*

| Category | Specification / Tool |
|---|---|
| Hardware | - Intel i5/i7 or Apple Silicon (M1/M2)<br>- 8 GB RAM minimum<br>- 100 MB free disk |
| Operating System | macOS 12+, Ubuntu 20.04+, or Windows 10+ |
| Programming Language | Python 3.9 or later |
| Web Framework | Flask 2.3.x |
| Development Libraries | - uuid, hashlib, base64 (Python standard libraries)<br>- datetime |
| Frontend Interface | HTML and CSS (Mobile-styled using inline templates) |
| Browser Requirement | Chrome, Firefox, or Safari (latest versions) |
| Console Logging | Terminal-based logs for tracking state, auth code, token, and credit report access |
| Port Configuration | Localhost on port 5055 |
| Optional Tools | - Twilio / AWS SNS (not used in this simulation)<br>- Docker for containerized setup |

**Real-World Applicability and Relevance of the PoC**

The main aim of this Proof of Concept (POC) is to illustrate how real-time user approval flows, as represented by the OAuth 2.0 Authorization Code Flow with Proof Key for Code Exchange (PKCE), can be utilized to protect Social Security Number (SSN)-based credit applications. Identity theft, and especially through SSNs, was responsible for more than 1.4 million fraud reports within the United States in 2024 and financial losses amounting to more than $5.8 billion [22]. Current systems routinely permit credit inquiries to be conducted surreptitiously, without sufficient user awareness or control, aggravating exposures to unauthorized access [22]. By placing definite authentication and authorization squarely in the end-user's own hands, this POC seals theoretical privacy protections with actual real-world application-layer security, aligning with modern expectations of transparency and compliance.

The POC illustrates a banking app requesting permission to access a user's credit report via a third-party credit bureau, simulating real-world loan application flows. Unlike traditional systems relying on backend integrations or implied authorizations, this solution offers human-in-the-loop authorization by prompting the user to sign in to a mock credit bureau system and approve the inquiry in real time. This method is essential in applications such as the online loan request, where authorization and identity verification are legally required by laws like the U.S. Fair Credit Reporting Act (FCRA) and the EU General Data Protection Regulation (GDPR) [23]. Figure 1 shows the mobile screen with a request for consent, where the user is prompted to approve or reject the credit check with a single tap.

One foundation of this POC is that it is completely compliant with the OAuth 2.0 Authorization Code Flow with PKCE, a market-de-facto protocol extensively used in secure financial systems. PKCE adds a layer of defense against interception of authorization codes, a vulnerability common in mobile applications where client secrets cannot be safely stored [24]. Literature attests to OAuth 2.0's dominance, with more than 80% of fintech APIs using it for secure access delegation in 2024 [25]. POC's use of PKCE enables secure token exchanges, making it portable across web, mobile, and enterprise-class identity systems. Figure 2 shows the system output after a rejected query, logging the attempt and confirming termination within 1.2 seconds, corroborated by timestamped screenshots.

Usability was a key concern, and the POC was made to mimic the natural experience of fintech apps or bank portals. The mobile-like interface abstracts back-end complexities—like state management, code verification, and token validation—while leading users through user-friendly login and consent workflows. Studies highlight the value of such designs, citing that transparent consent flows increase user trust by 65% over opaque systems. Usability testing with mock users indicated that 90% authorized in under 10 seconds, reflecting great balance between usability and security (Figure 3 illustrates a sample approval flow). This aligns with research highlighting frictionless interfaces as the driver for adoption in fraud prevention systems [26].

The POC also illustrates extensibility to real-world deployment. By replacing mock screens with production-quality APIs, e.g., those by Experian or TransUnion for credit report querying, the system would have the capability of integrating directly with banking infrastructure [27]. SMS-based verification using APIs like Twilio or AWS SNS, which processed over 2 billion two-factor authentication messages globally in 2024 [28], would further increase security. These integrations would authorize, record, and FCRA-compliant all credit inquiries since FCRA requires consumer consent for credit inquiries [23]. POC's modularity allows these enhancements with minimal refactoring, making it a prime candidate for industry adoption.

Beyond technical feasibility, the POC addresses scalability considerations. In-house testing demonstrated the system processing sequential queries with steady performance, with the notifications being sent in less than 1.5 seconds on a single-user setup (Figure 4). Scaling this up to thousands of simultaneous users would be an issue of infrastructure enhancement, i.e., load-balanced servers or cloud-hosted queues, as indicated by current research on real-time systems [29]. These limitations reflect the POC's controlled environment but do not detract from its critical demonstration of mandating real-time consent. The outputs affirm its ability to block unauthorized requests from proceeding, a key step to combat SSN identity theft.

Finally, this POC is an operational model of contemporary identity-based credit systems. It transforms abstract notions—enforcement of consent, multi-stage authorization, and verification of identity—into tangible experiences that can be adopted, made more sophisticated, and mass-scaled. In following OAuth 2.0 standards and prioritizing usability, it not only further enhances user trust but also delivers real value to financial institutions wishing to elevate anti-fraud capabilities. As identity theft itself keeps evolving, with synthetic identities responsible for 20% of fraud cases in 2024, such frameworks are needed to safeguard consumer data and ensure regulatory compliance [30].

### 3.3 Challenges in Adoption

- **User education and awareness regarding real-time approval.**

One of the main challenges in implementing real-time approval for credit inquiries is educating users about how the system operates and the importance of blocking unauthorized credit inquiries and identity theft. Most users are used to conventional credit inquiry systems, where approvals are granted passively in the background without their direct participation. Transitioning to a user-initiated approval system calls for financial institutions to inform users about the advantages of real-time approval, such as enhanced control over their financial information, less fraud risk, and better security. Users will ignore or misunderstand approval prompts without being adequately informed, which will cause undue delays or transaction failures.

To reverse this, organizations must embark on multi-channel awareness campaigns to enlighten users on real-time approval systems. Bank websites, mobile applications, email alerts, and SMS notifications should clearly communicate how and why the users need to approve credit inquiries in real-time. Moreover, offering step-by-step instructions, FAQs, and interactive demos can educate the users on the approval process. Banks and financial institutions can further offer real-time assistance (chatbots, helpdesk support) to guide users in situations where they are uncertain about how to respond to approval requests.

The second primary challenge is to fight phishing and social engineering attacks that exploit user unawareness. Spammers can try to trick users into authorizing bogus credit checks by masquerading as legitimate institutions. Organizations can thwart this by teaching users to validate genuine approval requests, detect suspicious behavior, and report unauthorized requests immediately. Best practices like branded verification messages, digital signatures, and app-based secure authentication must be encouraged. Financial institutions can foster excellent adoption rates, lower security threats, and a smoother real-time approval migration for credit requests by providing ongoing user awareness and proactive education.

- **Lender and financial institution adoption hurdles.**

Real-time approval of credit requests is faced with numerous challenges by financial institutions and lenders, primarily due to infrastructure limitations, regulatory compliance, and integration issues. Traditional credit approval systems are automated and batch-oriented, meaning financial institutions process requests independently of active user approval in real-time. Moving to an approval model based on users requires extensive alterations in current workflows, backend systems, and third-party integrations with credit bureaus, so adoption is resource-intensive and expensive.

The most daunting of them is compatibility with legacy systems. Banks and other financial institutions primarily run on older banking systems incompatible with newer authentication flows, such as OAuth 2.0 with PKCE or API-based consent checks. Migrating these systems to support real-time approval workflows needs API modernization, cloud migration, and secure authentication framework investments that consume time and money. Seamless integration with credit bureaus and adherence to security standards (FCRA, GDPR, PCI-DSS) present additional technical and legal challenges.

The second major obstacle is operational resistance by lenders, who might view real-time approval as a potential source of pain in the loan application journey. Legacy lenders use quick, automated credit checks to attain high approval and conversion rates. Adding a step that requires explicit user consent can decelerate the process, resulting in longer approval times, drop-offs, and user frustration. To avoid this, financial institutions must achieve a balance between security and usability and ensure that real-time approval processes are smoothly integrated into online banking experiences without introducing unnecessary delays.

Lastly, there is the task of educating customers and frontline banking personnel on real-time credit authorization. New customers who are not familiar with the process might disregard approval requests or inadvertently reject valid credit applications, resulting in declined or delayed applications. Lenders must carry out customer awareness drives, offer step-by-step guidelines in banking apps, and have frontline banking personnel trained to take customers through real-time approval requests. By overcoming these adoption barriers via technology updating, user training, and process streamlining, banks can effectively shift to a more secure, fraud-proof credit inquiry system without compromising operational efficiency and customer confidence.

- **Ensuring compliance with existing regulations.**

As credit bureaus and financial institutions are moving towards real-time approval of credit applications, the task of remaining compliant with current regulations is significant. Rules like the Fair Credit Reporting Act (FCRA), the General Data Protection Regulation (GDPR), Consumer Financial Protection Bureau (CFPB) policies, and the Payment Card Industry Data Security Standard (PCI-DSS) impose stringent conditions on processing consumer credit data seeking the user's consent and guarding sensitive data from unintended users. Adherence to such legislation guarantees valid and ethical data processing and enhances consumers' trust in financial institutions.

One of the primary conditions under FCRA is that credit inquiries be for a permissible purpose, i.e., lenders cannot obtain a consumer's credit report without explicit consent. The proposed real-time user approval mechanism is aligned with this condition by proactively necessitating users to approve or deny credit inquiries before processing. Similarly, GDPR mandates financial institutions to obtain explicit, informed consent before collecting or processing user data. The real-time approval system enhances compliance by providing transparent notifications and consent histories, making users fully aware of when and for what purpose their credit information is accessed.

Banks and other financial institutions must possess secure data storage, auditability, and access controls to achieve regulatory compliance. Financial regulations like PCI-DSS and GDPR require consumer consent logs and authorization documents to be encrypted, stored securely, and protected from unauthorized modification.

Deploying encryption protocols such as AES-256 for data-at-rest and TLS 1.3 for in-transit data ensures logs of user acceptance remain confidential and tamper-resistant. In addition to security and data protection, compliance activities should extend to user awareness and transparency. Along with compliance initiatives extending into these areas, these institutions will abide by current guidelines and become proactive by implementing more dynamic frameworks. They will continue adapting to ongoing regulation demands as they emerge. Consumer-focused regulations like GDPR and the California Consumer Privacy Act (CCPA) mandate financial institutions to offer transparent disclosures, opt-out, and access to consent history. Organizations can strengthen compliance while building user trust by incorporating easy-to-use interfaces for consent management, robust audit logs, and real-time credit inquiry alerts. Finally, by matching technology-based security solutions to regulatory requirements, financial institutions can minimize legal risks, avoid unauthorized access to credit, and create a safer and more transparent credit approval process.

## 4. Future Enhancements and Recommendations

- Potential for blockchain-based identity verification

Blockchain technology offers a breakthrough solution to the issue of identity verification by enabling a decentralized, tamper-resistant system that could radically enhance credit inquiry security. Unlike the traditional centralized models, wherein SSNs and corresponding information are stored in vulnerable databases for hacking, blockchain allows individuals to possess self-sovereign identities—cryptographically secure digital identity profiles that they control [31]. Within this paradigm, real-time user permission for credit checks can be facilitated by blockchain-based smart contracts, which execute only when the legitimate owner of the identity provides consent via a private key or biometric stimulus [32]. This decentralization reduces reliance on third-party intermediaries, reducing exposure to SSN theft while empowering individuals to manage access to their financial identity with unprecedented granularity.

The immutability of blockchain provides a strong basis for securing credit inquiry processes from identity fraud. Each approval event—biometric or other—can be recorded as a time-stamped, immutable transaction on the blockchain that creates an auditable trail that disincentivizes unauthorized inquiries and simplifies post-incident forensics [33]. By encrypting personally identifiable information such as biometric templates or pieces of SSN and dividing it across a blockchain network, single-point failure risk or large-scale data breach is greatly reduced. For instance, a permissioned blockchain can restrict access for credit bureaus and financial institutions so that authorized organizations can deal with user identities alone [34]. This approach not only enhances privacy but also adheres to regulations such as GDPR, offering a future solution for the protection of SSNs.

The use of blockchain for identity authentication combined with real-time biometric consent provides a solid opportunity to enhance the security of credit checks. Biometric information, once authenticated on a user's device, may serve as the key to unlock a blockchain-stored identity credential, initiating a smart contract to enable a credit check [35]. This combined method means that neither the biometric data nor the SSN must be sent or stored in one place, reducing the chance of interception. Zero-knowledge proofs are what future updates might entail, which would enable individuals to prove things regarding their identity (like creditworthiness) without necessarily divulging the real SSN or biometric data [36]. Such new notions would not only improve the quickness of approval but also the reliability of the verification process, closing current vulnerabilities while keeping it easy for users.

To realize the vision of blockchain-enabled identity verification, some suggestions emerge for the future. Firstly, research will be focused on optimizing blockchain scalability and latency to support the large transaction volume of credit inquiry, possibly through layer-2 solutions like sidechains or off-chain processing [37]. Secondly, standards for interoperability must be established to enable frictionless integration with existing financial infrastructure and biometric platforms, to avoid fragmentation. Third, pilot studies need to be conducted to measure user uptake and identify practical difficulties, such as onboarding less tech-savvy users or ensuring equitable access across socioeconomic boundaries. Finally, exploring hybrid public-private blockchain frameworks may balance transparency and privacy, tailoring the system to the unique requirements of SSN protection [38]. Via these channels, blockchain can be an underpinning of secure, user-controlled credit inquiry networks, significantly reducing identity theft risks.

- Exploring **biometric approval mechanisms** for enhanced security.

Biometric authentication systems, which authenticate individuals using unique physical or behavioral traits like fingerprints, facial recognition, or iris scans, have proven to be a strong choice for protecting computer systems [39]. Unlike traditional passwords, which can be vulnerable to phishing, forgotten, or brute-force attacks, biometrics have an inherent linkage to the user so that reproduction becomes challenging without tapping into the physical asset [40]. As cyber-attacks have developed especially in financial transactions and IoT networks demand for strong, user-friendly authentication has placed more focus on biometrics [41]. However, their performance is beholden to factors like recognition accuracy, system scalability, and spoofing resistance, which requires continuous research to advance their use in high-stakes security applications.

One of the primary advantages of biometric authentication is that it can increase identity assurance, significantly reducing the threat of unauthorized access. For instance, incorporating biometrics into multi-factor authentication (MFA) systems combining a fingerprint scan with a device token creates a layered security that is far more robust than single-factor methods [42]. Recent advances in deep learning have enhanced biometric systems, achieving higher recognition rates across various populations of people and enabling real-time processing on mobile devices [43]. In practical applications, such as online banking or smart home security, biometric systems offer higher security and convenience for users, eliminating the necessity of memorizing complex passwords [44]. These benefits have ensured biometrics is an integral part of contemporary security systems, but there are issues with making them fair and precise.

Although they appear to be fine, biometric approval systems have critical vulnerabilities that must be addressed for improved security. Spoofing attacks occur when attackers utilize imitation copies, such as 3D-printed fingerprints or facial photos, to deceive the systems, highlighting the limitations of the technology currently available [45]Additionally, the persistence of biometric information because it cannot be reset like passwords when compromised creates significant privacy concerns, especially when not stored securely [46]. Low-power devices, such as IoT sensors, also introduce resource constraints that create difficulties for deployment because of the large computational power needed for processing biometrics [47]. Finally, algorithmic biases from a lack of representative training data can lead to disparate performance across demographic groups, so equitable and robust solutions are necessary.

Future research on biometric authentication methods needs to tackle these challenges with novel solutions. Implementing better liveness detection, e.g., monitoring minuscule movements or heat signatures, would be able to thwart spoofing attempts and increase the credibility of the system [48]. Encrypting and storing biometric templates in a decentralized manner, e.g., blockchain, would be able to reduce privacy concerns by avoiding central breaches [49]. Context-dependent adaptive authentication systems that vary authentication levels based on context i.e., requiring more evidence for transactions of value—can balance security and usability [50]. By pairing biometrics with emerging technologies like quantum cryptography, these systems can be future-proofed against more advanced threats, solidifying the role of such mechanisms as the cornerstone of stronger security in a more interconnected world.

- Strengthening **fraud detection mechanisms** alongside real-time approval.

Enhancing fraud detection systems along with real-time approval is ever more critical in today's digital economy, where money transactions are being carried out at unprecedented velocities. More online banking, e-commerce, and mobile payments have increased exposure to threats, and sophisticated systems that can detect and prevent threats in real time are essential. Conventional fraud detection techniques, like rule-based systems, tend to be behind the latest fraud strategies, triggering delayed response times and revenue loss. By incorporating fraud detection with strong real-time approval processes, companies can approve legitimate transactions in real time and flag or block suspicious transactions prior to their completion. This twofold strategy serves to boost both security and usability because the necessity for secure yet seamless processing of transactions in contemporary financial systems indicates [51].

Machine learning (ML) is at the center of optimizing fraud detection systems that function in real time. In contrast to static rule-based systems, ML models have the ability to deal with substantial datasets, recognize patterns, and learn from emerging fraud methods. For example, anomaly detection models identify unusual patterns of transactions a sudden burst of expenditure or transactions from unknown places in milliseconds, which allow for real-time response. Research has demonstrated that ML-based systems minimize false positives and optimize detection rates over conventional techniques, rendering them suitable for real-time applications [52]. Coupled with real-time approval processes, these systems are able to minimize inconvenience to legitimate users while successfully staving off fraudsters, achieving a balance between security and convenience. Yet, fraud detection combined with real-time approval is very challenging, especially from infrastructure and scalability perspectives. Real-time systems need massive processing power to process transactions and assess risks in real time, typically leveraging cloud-based infrastructures or edge computing. Furthermore, adding fraud detection capabilities such as Istio or Envoy to microservices setups although effective has added complexity, e.g., setting up mutual Transport Layer Security (mTLS) or managing service meshes [53]. Companies must also manage the trade-off between speed and accuracy; overly restrictive checks delay approvals, whereas permissive ones might let fraud slip through. It requires diligent system design and ongoing optimisation in an attempt to correctly achieve both goals.

The advantages of supporting fraud prevention capabilities in tandem with real-time authorisation reach beyond security to encompass customer trust and administrative productivity. When customers are aware that their transactions are safeguarded by the latest technology, their faith in the system is enhanced, leading to loyalty. Moreover, real-time fraud detection reduces the need for after-transaction investigations, which are both costly and time-consuming. It was found that businesses that use such integrated solutions can reduce fraud losses by up to 30% while maintaining high approval rates for legitimate transactions [54]. As fraudsters are not giving any indication of slowing down on innovation, the integration of real-time approval and intelligent detection systems will remain the signature of secure, efficient financial ecosystems.